\documentclass[ a4paper, 10pt,twocolumn]{article}
\usepackage{graphicx}
\usepackage{cite}
\title{Spatial Light Modulators for the Manipulation of  Individual Atoms}

\author{Lukas Brandt, Cecilia Muldoon, Tobias Thiele, Jian Dong, \\ Edouard Brainis and Axel Kuhn\\
Clarendon Laboratory, University of Oxford, \\Parks Road, Oxford, OX1 3PU, UK; 
\\e-mail: a.kuhn1@physics.ox.ac.uk }

\begin{document}

\maketitle

\begin{abstract}
We propose a novel dipole trapping scheme using spatial light modulators (SLM) for the manipulation of individual atoms. The scheme uses a high numerical aperture  microscope to map the intensity distribution of a SLM onto a cloud of cold atoms. The regions of high intensity act as optical dipole force traps. With a SLM fast enough to modify the trapping potential in real time, this technique is well suited for the  controlled addressing and manipulation of arbitrarily selected atoms.
\end{abstract}

\section{Motivation}

Many proposals for large-scale quantum computing \cite{vincenzo1} or quantum simulation \cite{qsimulation} rely on the ability to deterministically manipulate, address and couple individual components of a quantum network. These challenges can be adressed by the use of isolated quantum systems, such as trapped ions in quantum computing and entanglement experiments\cite{qipions1, qipions2},  or dipole-trapped single neutral atoms\cite{meschede2, grangieratommovement, bloch2}. In an effective one-dimensional arrangement it has already been shown that optical addressing and preparation of individual dipole-trapped atoms is possible and that those atoms can also be used as a phase preserving quantum register\cite{schrader}. To control a larger number of atoms in a two or three-dimensional quantum network, more elaborate techniques are required. Therefore several methods for handling individual atoms have been developed. The most promising techniques are the trapping of ion strings\cite{qipions1, qipions2, ionstring}, magnetically trapped neutral atoms above atom chips \cite{hinds,schmiedmeyerneu,hansel}, individual atoms in steep optical dipole-force traps\cite{grangieratommovement}, optical lattices\cite{bloch2},  and  dipole-trap arrays created either by a matrix of lenses\cite{birkl2}, by  holograms\cite{grangierholographic}, or by a combination of these techniques\cite{atomomicroscope}. In contrast to these attempts, we propose to use a spatial light modulator (SLM)\cite{macaulay, hanley} to form the desired trapping potential.  This concept  has much in common with the optical tweezers technique used in biology \cite{opticaltweezersvirus}, where light is used to manipulate small objects, like viruses, bacteria, or organic samples attached to micro-spheres. Here we propose to apply this technique directly to individual atoms.  This combination of the SLM with  optical tweezers offers atom-chip like flexibility of individual-atom manipulation as well as magnetic-field free dipole trapping. Furthermore, with SLM refresh rates far above the loss rate of  trapped atoms, such an arrangement might allow for the real-time manipulation of atoms.

\section{Outline}

\begin{figure}[!h]
\includegraphics[height=7cm]{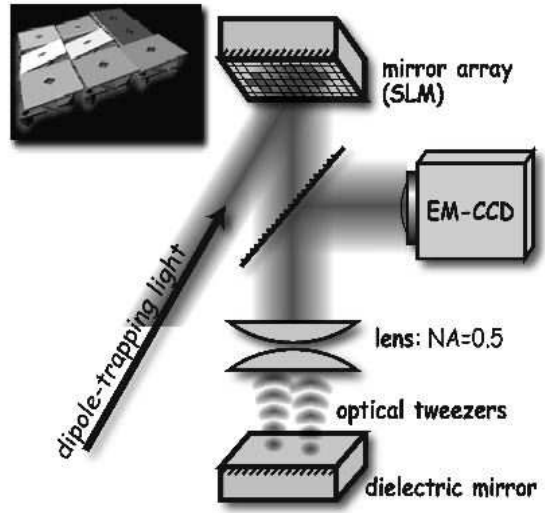}
\caption{Scheme of the proposed optical tweezers. A red detuned laser beam illuminates the mirror array of the SLM. A microscope reproduces the intensity pattern of the SLM at the surface of a mirror placed in the tweezer plane (object plane of the microscope). The reflected light interferes with the  incoming beam and forms a standing wave.  Its antinodes act  as dipole traps, which can hold individual atoms. Since the SLM is reconfigurable the setup allows arbitrary repositioning of atoms.  A very sensitive CCD camera, detecting atomic fluorescence via a dichroic mirror, can be used to image the trapped atoms and monitor the tweezers in operation.  The insert (Courtesy of Texas Instruments) shows a block of 3 by 3 mirrors of the SLM. The mirrors can be switched between two tilt angles, so that the light is  directed either into the microscope or deflected  to a beam stop.}
\label{tweezers}
\end{figure}

At the heart of this novel scheme is the use of a SLM to generate arbitrary patterns of light in the object plane of a microscope and to trap atoms therein. Fig.\,\ref{tweezers} illustrates this approach and shows how one can use the same microscope to observe the atoms. Several types of  SLM are  available which all have the potential to achieve this goal. One widely used technology is the liquid crystal device (LCD) \cite{grangierholographic, LCDBirkl}. It has the advantage of creating a range of grey scales, but has the drawback of having a low refresh rate, usually not exceeding 100\,Hz. This is in general too slow for the dynamic control of  trapped neutral atoms. Other commercially available  SLMs are digital mirror devices (DMD) \cite{TIpaperonDMD}. They consist of an array of individual flat mirrors which can be independently  switched between two tilt angles to generate arbitrary intensity distributions. With a full-frame refresh rate of up to  50\,kHz and a resolution of typically 1024 by 768 individual mirrors, these devices seem ideal for the real-time manipulation of trapped atoms.  

\begin{figure}[t!h]
\includegraphics[height=6.9cm]{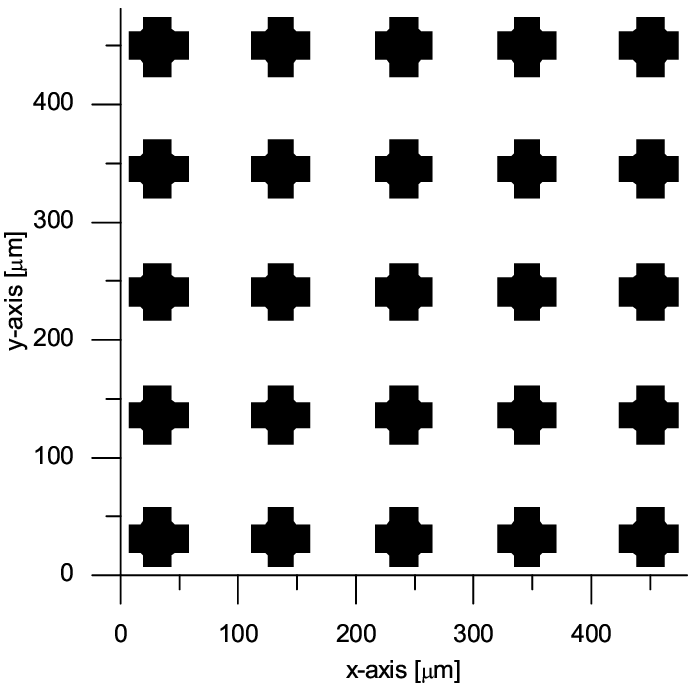}
\includegraphics[height=6.9cm]{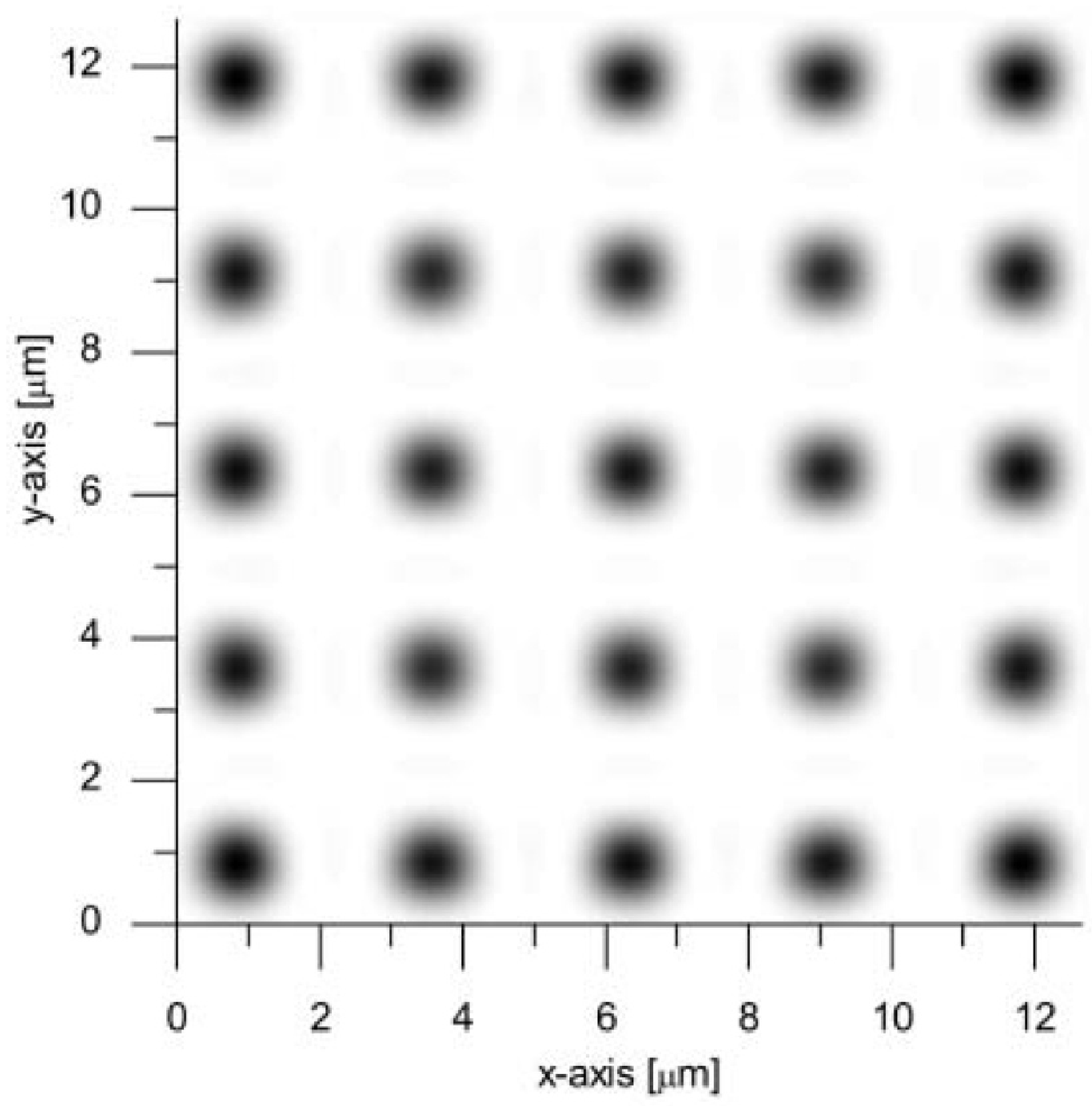}
\caption{The top picture shows a pattern on the SLM. The pattern is a repetition of a 4 by 4 mirror block where the corner pixels are missing.  Black marks the mirrors which are switched on.  Each mirror has a dimension of 14$\,\mu$m by 14$\,\mu$m. The bottom picture shows the resulting image, where black is the highest intensity, when an imaging system with a numerical aperture of $N\!A=0.5$ and a magnification of 38 is used. }
\label{pattern}
\end{figure}

The desired intensity distribution  can be formed in two ways. 
One possibility  is to place the SLM in the back focal plane of a lens system, such that its Fourier transform is in the front focal plane. Thus the SLM acts as an effective hologram for the resulting intensity distribution \cite{grangierholographic}. 
This method has the drawback that any change in the trapping pattern requires one to recalculate the entire hologram and thus restrains the experimenter to work with predefined sequences. Alternatively, the tweezers can be generated by imaging the surface of an \emph{amplitude-modulating} SLM through a microscope directly onto the atoms. The tweezers are then formed in the object plane of the microscope, see  Fig.\,\ref{pattern}. In the following we concentrate on this latter method, as the direct mapping of intensities provides an intrinsically higher speed and flexibility in the manipulation of trapped atoms. This system can be used to expose the atoms to almost any arbitrarily shaped and time-varying potential landscape, which offers a plethora of applications. Out of these we focus on the implementation of an array of very tiny atom traps, eventually capable of storing single atoms.  

Any tight spatial confinement of atoms  in dipole-force traps relies on small foci of the trapping light. Therefore, an optical system of high numerical aperture is necessary. This can be done with either a sophisticated microscope objective \cite{lensbonn} or a single aspheric lens\cite{aspgrangier}. Typically, these are diffraction limited and have a numerical aperture of 0.5,  leading to an optical  resolution approximately equal to one wavelength, $\lambda$. This also determines the lateral confinement of the atoms. Along the optical axis, the atoms are trapped by the longitudinal beam profile. For a single beam focussed to $\lambda$,  atoms are axially confined to a Rayleigh length of $\pi\cdot\lambda$. To obtain a better confinement along the optical axis, a standing wave could be used.  This can be achieved  by placing a mirror at or close to the plane where the tweezers are formed. The atoms trapped in the anti-nodes of the standing wave are then axially confined to $\lambda/2$. 

The underlying physical mechanism used to  trap atoms in optical tweezers is the optical dipole force, which atoms experience in detuned light fields. This force is conservative but  very weak, resulting in a shallow potential with a typical depth of  $\sim 1$\,mK. Only atoms with  low kinetic energy can be trapped,  and therefore they need to be pre-cooled before loading them into the tweezers. This is usually  accomplished by a magneto-optical trap (MOT), trapping and cooling atoms down to the Doppler cooling limit, $T_D\sim 100\, \mu \textnormal K$ for most alkali atoms \cite{motalkalie}. In the case where a mirror is included to form a standing-wave dipole trap, the pre-cooling has to be done close to the mirror surface. This can easily be accomplished using a magneto-optical surface trap (MOST)\cite{most-paper}.

\section{Feasibility} 
\subsection{Dipole trapping of Alkali Atoms}
The majority of experiments  cooling and trapping neutral atoms are done with alkali atoms, of which Rubidium is one of the most prominent species.  Hence we choose it for the purpose of this feasibility study. The optical tweezers consist of tiny optical  dipole-force traps, in which the dynamic Stark shift gives rise to a trapping potential in the \emph{far} detuned limit  \cite{chrisfoot}:

\begin{equation}
U_{dip}(\mathbf{r})\approx\frac{\hbar\Gamma^2 }{8\delta}\frac{I(\mathbf{r})}{I_{sat}}
\label{eq1}
\end{equation}

\noindent where  $\delta$ is the detuning of the trapping laser light with respect to the atomic transition, $I(\mathbf{r})$ is the intensity of the trapping light, $I_{sat}$  is the saturation intensity for the chosen transition, and $\Gamma$ is the corresponding decay rate. For red detuned light, $\delta < 0$,  $U_{dip}$ is negative and hence attractive. The most critical parameter is the trap depth $U_0$, which corresponds to the highest intensity of the trapping laser light $I_0$.  Hence $U_0$ is the energy required for an atom at rest to escape the trap, neglecting gravity. Obviously, the trap depth has to be at least as large as the energy of the pre-cooled atoms.  Therefore, the minimum usable trap depth is determined by the temperature that can be reached by the magneto-optical trap. For rubidium, the Doppler cooling limit would be $T_{D}= 143 \,\mu\textnormal{K}$. 
However, it is desirable to operate deeper traps since the spatial confinement of the atoms increases with depth. For atoms whose kinetic energy is smaller than the potential depth by a sufficient amount, the trap can be treated as harmonic, and the spatial confinement of the atoms can be determined by the harmonic oscillation frequencies 
$\omega_r$ and $\omega_z$ for radial and longitudinal motion, respectively. The achievable values of $U_0$, $\omega_r$, and $\omega_z$ depend on the wavelength $\lambda$ of the trapping laser, its intensity, and the trap geometry.

\begin{figure}[!ht]
\includegraphics[height=09cm]{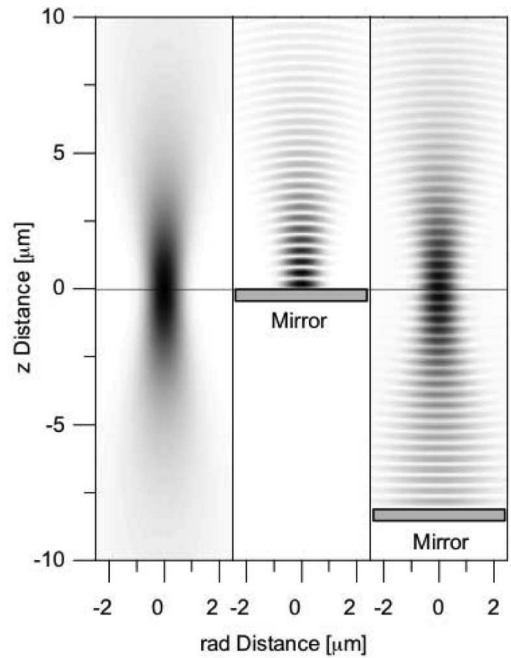}

\caption{Intensity distribution of a focused laser beam (with beam waist $w_0=0.78\,\mu\textnormal{m}$) in the neighborhood of the focal plane for different distances $L$ between the dielectric mirror and the focal plane. (Black corresponds to the highest intensities.) The left picture is for $L\rightarrow \infty$, i.e. no mirror being present. The tweezer is a pure forward traveling wave with no longitudinal intensity modulation. The middle picture shows a beam which is reflected in its focal plane ($L=0$). The tweezer has a perfect standing wave pattern, but atoms cannot be trapped in the beam waist. The right picture shows the intensity of a retro-reflected beam whose focus lies $L=8 \mu$m above the mirror surface. Here the intensity modulation is notably less than in the previous case.}
\label{stwa}
\end{figure}

In the proposed scheme (Fig.~\ref{tweezers}), the trap geometry is essentially determined by the distance $L$ between the focal plane of the optical system which generates the tweezers and the dielectric mirror. The intensity modulation induced by the interference of forward and backward travelling waves over this distance can be seen in Fig. \ref{stwa}. In the following, we first discuss the limiting case $L\rightarrow\infty$ (left picture of Fig. \ref{stwa}), which corresponds to having no mirror at all, and  then analyze the longitudinal intensity modulation for finite $L$.

When the dielectric mirror is not present ($L\rightarrow\infty$), the trapping potential has a single minimum at the beam waist, in the focal plane of the lens system. The maximum achievable intensity at that point,
\begin{equation}
I_{0} = I_{d} \cdot M^2 \label{eq2},
\end{equation}
depends of the magnification $M$ of the optical system and the damage threshold the DMD with respect to the incident intensity, which is about $I_d=10$ W/cm$^2$. To maximize the spatial control of the atoms, the individual traps should be as small as possible. The smallest achievable trap size is set by the resolution of the optical system $D_{min}$. According to the Raylight criteria it is: 
\begin{equation}
D_{min}=1.22\cdot\lambda N,\label{diflim}
\end{equation}
where $\lambda$ is the wavelength of the light used, and $N$ is the f-number related to the numerical aperture $N\!A$ of the optical system by $N\approx \frac1{2 N\!A}$. If  an aspherical singlet lens is used,  the $N\!A$ can be as high as 0.5. For this value, the resolution limit is given by about the wavelength of the trapping light, which is $\lambda\approx 780\,\textnormal{nm}$  for the D$_2$ line of Rb. 

\subsection{The role of the SLM}

\begin{figure}[!ht]
\includegraphics[height=5.05cm]{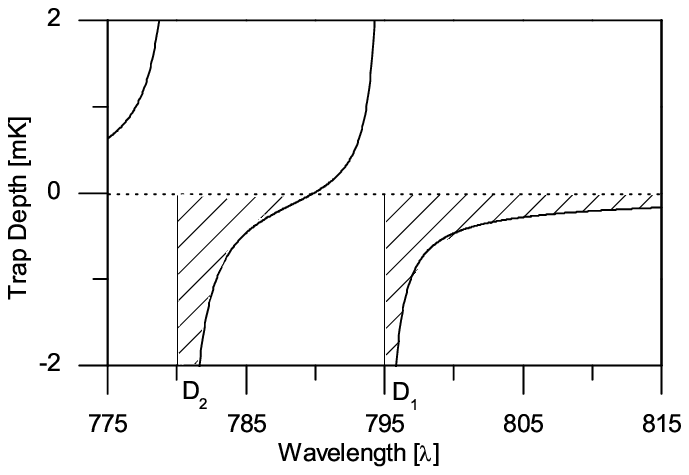}
\includegraphics[height=5.05cm]{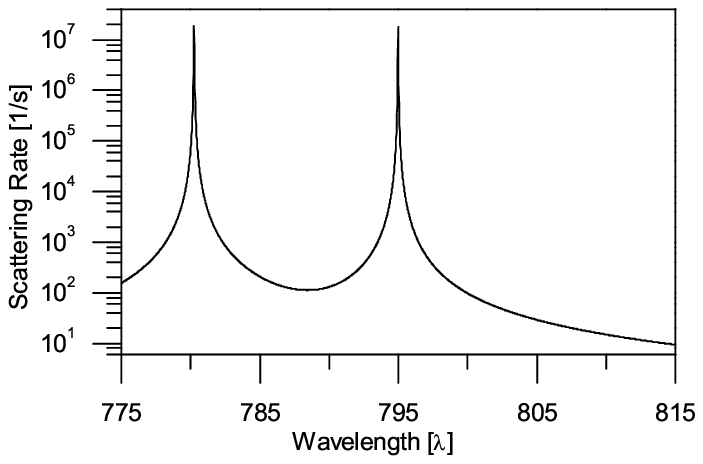}
\caption{The top graph shows the potential depth of the trap with respect to the wavelength of the trapping light, for linearly polarized light with an intensity of $I_0=14.4\,\mathrm{kW/cm}^2$. The trapping regions are hatched. The bottom graph shows the corresponding scattering rate.  }\label{f2mf0}
\end{figure}

It is important to bear in mind that any experiments involving dynamics require the reconfiguration of the trapping potential. One way atomic transport could be realized is by displacing a pattern of traps. The switching of mirrors would lead to abrupt changes in  the trapping potential if the contribution of light for a point in the image plane of the DMD is dominated by one mirror only. If, however, the optical system does not resolve individual mirrors, the reconfiguration of the traps can be smoother, since the light intensity for every point will have contributions from a number of mirrors.  Effectively, this  leads to  grey scales in the trapping pattern.    To take advantage of this, we choose the magnification $M$ such that blocks of $2 \times 2$ mirrors will just about be resolved. The micro-mirrors of a commercially available DMD have a size of $14\,\mu\mathrm{m} \times 14 \mu \mathrm{m}$ each, so in our case,  a  magnification of $M = 38$ is required for a diffraction limit of  $D_{min} \approx  780\, \mathrm{nm}$ (see Fig. \ref{pattern}).
According to Eq. (\ref{eq2}), if the DMD is illuminated with light of an intensity at  its damage threshold, an intensity $I_0$ of $14.4\,\mathrm{kW/cm}^2$ can be obtained at the center of a tweezer.
 The saturation intensity for the D$_2$ line of Rubidium is 
 $I_{sat}= 2.5\,\mathrm{mW/cm}^2$. Therefore the maximum intensity is $5.8 \cdot 10^6$ times the saturation intensity. Taking the above discussion into account and requiring  a trap depth of $1\ \mathrm{mK}$ leads to a detuning $\delta \leq 2.1\cdot 10^5 \Gamma$, according to Eq. (\ref{eq1}). This corresponds to a wavelength detuning of about $2.6
\ \mathrm{nm}$ to the red with respect to the D$_2$ line. However, since Rubidium has a fine structure splitting of $15\ \mathrm{nm}$,  this estimation of the required detuning is certainly too crude. The hyperfine transitions within the D$_1$ and D$_2$ lines of Rb need to be taken into account. Two possible wavelength regimes present themselves. The first is slightly red detuned from the D$_2$ line at $780\ \mathrm{nm}$, and the second is red detuned from the D$_1$ line at $795\ \mathrm{nm}$ (see the top graph of Fig. \ref{f2mf0}).  More specifically, for a trapping laser  with  an intensity of $I_0=14.4\,\mathrm{kW/cm}^2$, for instance, a trap depth of 1$\,$mK can be achieved for the wavelengths of $\lambda_2=782.85\ \mathrm{nm}$, and $\lambda_1=796.90\ \mathrm{nm}$. 

Working with light so close to the D$_1$ and D$_2$ resonnances gives rise to atomic heating due to photon scattering from the trapping beam. The heating is proportional to the scattering rate $R_{scat}$, which,  in  the far detuned limit\cite{chrisfoot}, is given by:

\begin{equation}
R_{scat}\approx\frac{\Gamma^3 }{8\delta^2}\frac{I}{I_{sat}}.\label{scattering}
\end{equation}

\noindent Here we have to take  the scattering from both the D1 and the D2 line into account, as  shown in the bottom graph of Fig. \ref{f2mf0}.  For both $\lambda_1$ and $\lambda_2$, Rb scatters about 600 photons per second. The scattering rate in both detunings is about the same,  since in both cases, the detuning to the closest transition is  small when compared to detuning from the  other transition. If  the trapping laser were further detuned, the region to  the red from the D$_1$  line would  be preferable, as the potentials of both lines add constructively there, whereas in the region in-between the D$_1$ and D$_2$ lines, the potentials add destructively. 

Given that each scattering event increases the kinetic energy of an atom by an amount equal to the recoil energy, the lifetime of a Rubidium atom (initially at rest) in the dipole trap would be 2.3$\,$s if photon scattering was the only source of heating. It will be shown below that this time is long enough to allow for many transport and detection cycles.

\subsection{Radial and axial confinement}

We now estimate the radial and longitudinal confinement for a travelling-wave dipole trap ($L\rightarrow \infty)$. The intensity of the light field in the radial direction is  
\begin{equation}
I(q)=I_0\left( \frac{2\ J_1(q)}{q}\right)^2,\label{intensitybessel}
\end{equation}
 where $q=\frac{\pi r}{\lambda N}$, $r$ is the distance from the axis of the trap, and $N\approx 1/(2\,N\!A)$ is the f-number which is about 1 in this case. $J_1(q)$ is a Bessel function of the first kind. For small $q$, it expands as
 \begin{equation}
J_1(q)=\frac{1}{2} q- \frac{1}{16} q^3+O(q^5)\label{besselex}.
\end{equation}
Inserting this  into  (\ref{intensitybessel}) and keeping only terms up to second order we obtain the harmonic approximation to the trapping potential in the radial direction 
\begin{equation}
U(r)\approx U_0 -\frac{U_0}4 \left( \frac{\pi}{\lambda N} \right)^2 \ r^2,
\end{equation}
where $U_0$ is the trap depth. It is plotted in Fig. \ref{f2mf0} as a function of the wavelength $\lambda$. A particle with mass $m$ oscillates with frequency
\begin{equation}
\omega_r=\sqrt{\frac{-U_0\pi^2}{2m\lambda^2N^2}}
\end{equation}
in this potential. For $-U_0= 1\ \mathrm{mK}\cdot {k_B}$, this leads to a radial trapping frequency of
 $\omega_r=2\pi \cdot  162\,  \mathrm{kHz}$.
If we approximate this as a Gaussian distribution, the beam waist  would be $w_0=\sqrt{2} \lambda N/\pi=0.61 \ \mu$m,  which is on the same order of magnitude as the resolution limit determined with (\ref{diflim}). The oscillation frequency along the axial direction $\omega_z$ can be calculated using the Rayleigh length $z_0=\pi \ w_0^2/\lambda\approx 1.5 \ \mu$m:
\begin{equation}
\omega_z=\sqrt{\frac{-2U_0}{mz_0^2}}.
\end{equation}
For $-U_0=1\ \mathrm{mK}\cdot {k_B}$, this yields  $\omega_z=2\pi\cdot 46\,  \mathrm{kHz}$. The axial trapping frequency $\omega_z$ is significantly lower than the radial trapping frequency $\omega_r$, which is generally the case for running wave optical dipole-force traps. 
 
\begin{figure}[!h]
\includegraphics[height=5cm]{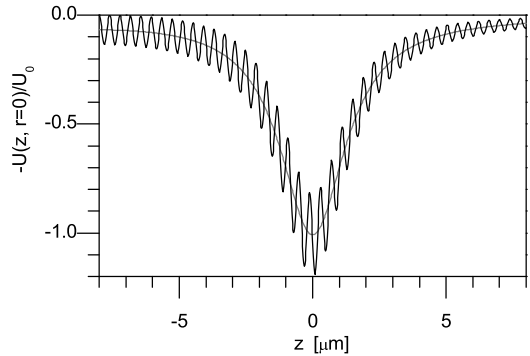}
\caption{Potential depth U(z) along the axis, with the focal point 8$\,\mu$m away from the mirror surface.  The slowly varying  average of the potential (gray) is a feature of focusing the light, whereas the oscillation (black) is caused by the partial standing wave. The modulation has the highest visibility close to the mirror-surface, becomes less pronounced in the focal spot, and then increases again further away from the focal point. }
\label{partialcross}
\end{figure}

We now consider the dielectric mirror to be in the object plane of the microscope $L = 0$. The backward travelling wave interferes with the forward travelling wave and
gives rise to a longitudinal interference pattern, creating
an ideal standing wave (see the central picture of Fig. 3).
The potential has many local minima in which atoms
can be trapped. The intensity modulation
along the z-axis gives rise to the following longitudinal
potential:
\begin{equation}\label{st}
U(z)=A(z) \sin^2\left(\frac{2\pi}{\lambda} z\right).
\end{equation}
The envelope of the modulation, $A(z) =4U_0/(1 + z^2/z_0^2)$,  reaches its maximum at the mirror surface. Here $U_0$ is the potential depth only due to to a single focused beam. Because the contrast of the modulation is one, $A(z)$ is also the trap depth of a local minimum at height $z$ above the mirror. The largest achievable trap depth is $4U_0$, four times higher than without mirror. Therefore 
the maximum radial trapping frequency in the standing wave configuration
is $\omega^{st}_r=2\omega_r=2\pi \cdot 324$ kHz and is achieved close to the mirror.
The longitudinal oscillation frequency inside a local trap at height $z$ is given by 
\begin{equation}\label{freqz}
\omega^{st}_z(z)=\sqrt{\frac{-8 A(z) \pi^2}{m\lambda^2}}.
\end{equation}
Close to the mirror, the highest value $\omega^{st}_z=2\pi \cdot 1100$ kHz is achieved.

When the mirror is not exactly in the image plane of
the DMD (finite distance $L$), the forward and backward propagating waves do not have the same amplitude. Therefore the longitudinal intensity
modulation, whose peak-to-peak amplitude is 
\begin{equation}\label{A}
A(z)=\frac{4U_0}{\sqrt{1+z^2/z_0^2}\sqrt{1+(2L-z)^2/z_0^2}},
\end{equation}
rapidly fades away as the distance between the mirror and the focal plane increases. Fig. 5 shows the longitudinal potential $U(z)$ for $L = 8$ $\mu$m. The deepest local minima are found at the focus of the beam. An atom trapped in one of these local minima sees a trap depth of about $1.37~U_0$ in the radial direction but only $0.3~U_0$ in the longitudinal direction. Whatever the distance between the focal plane and the mirror, the trap depth in the radial direction will always be larger than in the travelling wave case ($L\rightarrow \infty$) but smaller than in the standing wave case ($L=0$): $162~\mathrm{kHz}<\omega_r/(2\pi)<324~\mathrm{kHz}$. Along the z-axis, the trapped atoms are confined either to the slowly varying envelope or to the local minimum, depending on their initial kinetic energy. Only the coldest atoms can be trapped in the local minima. If they are, their confinement shall be strongly improved. In that case the longitudinal frequency can be estimated from Eq. \ref{freqz}, with $A(z)$ given by Eq. \ref{A}. For the example of Fig. 5, one finds that the oscillation frequency at the focus of the beam is $\omega_z=2\pi \cdot 300$ kHz. Depending on $L$, the values of $\omega_z$ are intermediate between the travelling wave case ($L\rightarrow \infty$) and standing wave ($L=0$) case:  $46~\mathrm{kHz}<\omega_z/(2\pi)<1100~\mathrm{kHz}$.

\section{Ways to single atoms}

In addition to controlling the size and depth of the traps, it is also important to control and measure the number of atoms. More specifically for the goal of conducting single atom experiments an efficient scheme to prepare single trapped atoms is essential. When loading atoms into moderately sized traps, the number of atoms therein is generally Poisson distributed. In this case, we  would need to measure the atom number and then post-select traps with just one atom. Fortunately, for tightly focused dipole traps, the collisional blockade mechanism \cite{grangierblock} should give rise to a significant departure from the Poissonian statistics and favor the loading of only one atom per trap. 

The effect occurs when atoms are loaded at a rate $R$  from a magneto optical trap into tightly confined dipole traps. It relies on light-assisted  collisions \cite{loadingadipoletrap} between atoms in the presence of the MOT cooling beam. Provided the latter is strong enough to saturate the transition, two-body losses occur at a rate   $\beta  ' N(N-1)$. $N$ is the atom number and $\beta  '$  is the rate constant, which is inversely proportional to the trapping volume. 
There is also a single body loss process $\gamma N$, which is mostly due to background collisions, so the full rate-equation\cite{schlosser2}  for the entire loading process reads:

\begin{equation}
\frac{dN}{dt}=R-\gamma N -\beta  ' N(N-1).
\end{equation}

Collisional blockade will  occur  when the two-body loss rate dominates over the loading rate of the dipole trap. Hence, it will work very efficiently when $\beta '$ is large, i.e. if the atoms are confined to a tiny volume. The volume occupied by the atoms depends on the temperature and the trapping frequency.   A comparison of the frequencies achievable with the proposed set up, and the trapping frequency $\omega_t =2\pi\cdot 200 \,\textnormal{kHz}$ of a previous experiment demonstrating the collisional blockade \cite{grangierblock} leads us to the conclusion that this mechanism could here be highly efficient as well.

\section{Single atom detection and transport}

To investigate the loading processes, the atom numbers in the individual traps have to be determined. For well-isolated traps, laser induced resonance fluorescence could be used for atom counting\cite{atomcounting}. To collect the fluorescence, the lens system which creates the optical tweezers  can be used in the backward direction as a microscope.  This fluorescence is detected by a highly sensitive camera. 

To discuss  the feasibility of single atom detection, the flux of photons impinging on the camera has to be determined.  The geometrical collection efficiency $\eta$ 
  depends on  the numerical aperture of the lens system. To evaluate $\eta$, we assume a uniform photon emission of the atoms and  calculate the ratio of photons passing through the system. The ratio of the surface area of a spherical cap with that of a sphere, yields 
   $\eta=2\pi\left(1-\cos(\frac{\alpha}2) \right)/(4\pi)$,  where $\alpha$ is the opening angle. For a numerical aperture of 0.5,  $\alpha=60\,^{\circ}$,  we expect  a collection efficiency of $\eta=0.067$.

If the radiative transition of the atom is completely saturated, the scattering rate on resonance tends to  $R_{scat}= \frac12\Gamma$, 
hence $R_{scat}=19.6\, \mu \textnormal{s} ^{-1}$ for Rubidium. 
  Assuming a further loss of 50\% through the lens system, a camera should collect photons from  a single atom with a rate of  $R_{cam}=653\,\textnormal{ms}^{-1}$. 
  In other words, a camera operating with an exposure time of $100\, \mu\textnormal{s}$ could count as many as 65 photons. For modern electron multiplying charged coupled device  (EM-CCD) cameras, this is well above the signal to noise threshold, so it is possible to use such devices to observe single atoms.

The high frame rate of the DMD should  allow for  the reconfiguration of the potential 
landscape, and hence for the transport of confined atoms. A naive method would be to transport atoms 
adiabatically by changing the potential slowly enough so that the atoms are following without  being heated. Since the DMD has only a finite resolution, this must be done using discrete steps. With the aforementioned  trap consisting of a four by four block, as shown in  Fig. \ref{pattern},
the pattern would have to be  shifted by at most one pixel per step.  The longest distance the traps could be moved is 1000 pixels, that is from one edge of the DMD to the other. With a refresh rate of $50\,\textnormal{kHz}$, this transport would take $20\,\textnormal{ms}$. This is much shorter than the anticipated lifetime of the trap, which should be of at least several $100\,\textnormal{ms}$. However a drawback of this method is that discrete switching of mirror elements leads to parametric heating of the atoms. To circumvent this,  the atom-transport could  also be realized ballistically. The initially trapped atoms would be accelerated through a sudden change in trapping potential,  guided through a channel and eventually decelerated and recaptured. This scheme has the advantage that it needs only a few changes in the potential, making it less susceptible to heating, and yet faster at the same time. 

\section{Conclusion and outlook}

Our study shows that the handling and manipulation of individual dipole-trapped atoms should be possible with an optical-tweezers type setup with the technologies at our disposal today. 
Using this new scheme, real-time arrangement of dipole traps is possible, and  two-dimensional transport of single atoms seems feasible. We expect this technique to boost the field  of  quantum computing, as it will allow to handle single entities within a large scalable array. For instance pairwise entanglement through either cavity QED \cite{cavityentanglement}, or controlled collisions \cite{controlledcollisions} could be used for generating large cluster states which are the essential resource for one way quantum computing\cite{oneway} .

\section{Acknowledgement}
 We  acknowledge the support from  the Engineering 
and Physical Sciences Research Council (EP/ E023568/1), the Research Unit 635 of the 
German Research Foundation,  the Belgian Fonds de la Recherche Scientifique - FNRS, the Philippe Wiener and Maurice Anspach Foundation, and the EU through the research and training network 
EMALI (MRTN-CT-2006-035369) and the integrated project SCALA. We are also grateful to Matt Himsworth for helpful discussions.

\end{document}